\documentclass[aip, pop, graphicx, amsmath, floatfix, reprint]{revtex4-1}
\usepackage{graphicx}
\usepackage{lipsum}
\usepackage{array}
\newcolumntype{C}[1]{>{\centering\let\newline\\\arraybackslash\hspace{0pt}}m{#1}}

\begin{document}


\title{Investigation of the Effect of Resistivity on Scrape Off Layer Filaments using Three Dimensional Simulations} 



\author{L. Easy}
\email[]{le590@york.ac.uk}
\affiliation{Department of Physics, University of York, Heslington, York, YO10 5DD, UK}
\affiliation{CCFE, Culham Science Centre, Abingdon, OX14 3DB, UK}

\author{F. Militello}
\affiliation{CCFE, Culham Science Centre, Abingdon, OX14 3DB, UK}

\author{J. Omotani}
\affiliation{CCFE, Culham Science Centre, Abingdon, OX14 3DB, UK}

\author{N.R. Walkden}
\affiliation{CCFE, Culham Science Centre, Abingdon, OX14 3DB, UK}

\author{B. Dudson}
\affiliation{Department of Physics, University of York, Heslington, York, YO10 5DD, UK}


\date{\today}

\begin{abstract}
The propagation of filaments in the Scrape Off Layer (SOL) of tokamaks largely determine the plasma profiles in the region.  In a conduction limited SOL, parallel temperature gradients are expected, such that the resistance to parallel currents is greater at the target than further upstream.  Since the perpendicular motion of an isolated filament is largely determined by balance of currents that flow through it, this may be expected to affect filament transport.  3D simulations have thus been used to study the influence of enhanced parallel resistivity on the dynamics of filaments.  Filaments with the smallest perpendicular length scales, which were inertially limited at low resistivity (meaning that polarization rather than parallel currents determine their radial velocities), were unaffected by resistivity.  For larger filaments, faster velocities were produced at higher resistivities, due to two mechanisms.  Firstly parallel currents were reduced and polarization currents were enhanced, meaning that the inertial regime extended to larger filaments, and secondly a potential difference formed along the parallel direction so that higher potentials were produced in the region of the filament for the same amount of current to flow into the sheath.   These results indicate that broader SOL profiles could be produced at higher resistivities. 
\end{abstract}


\maketitle 

\section{Introduction} 
\label{sec:intro}
One of the biggest challenges facing future generations of magnetic confinement devices such as ITER is to control the high particle and heat fluxes at the divertor surfaces.  These fluxes are determined by the balance between transport across and parallel to the magnetic field in the Scrape Off Layer (SOL), with enhanced cross field transport leading to a broader SOL width and hence reduced fluxes to the divertor.  

Perpendicular transport of particles (and to a lesser extent heat\cite{LaBombard:2001ks}) in the SOL has been observed to be dominated by the radial advection of coherent plasma structures \cite{Boedo:2003fe}, that are significantly more dense and hot than their surrounding plasma, with peak fluctuations typically of the order of the background \cite{Zweben:2007dp}.  These structures are aligned to the equilibrium magnetic field and strongly localized in the drift-plane perpendicular to it and hence are referred to as \textit{filaments} (or \textit{blobs} due to their appearance in the perpendicular plane).  A number of recent works have provided reviews of the experimental evidence for filaments\cite{Zweben:2007dp}, the theoretical understanding of their propagation and contribution to SOL transport \cite{Krasheninnikov:2008fw, Garcia:2009fs}, and of the agreement found when experimental measurements of filaments are compared with theory and simulations\cite{DIppolito:2011di}.  

The basic mechanism by which filaments advect radially outwards from the core was first proposed by Krasheninnikov \cite{Krashenninikov:2001uc} and can be understood in a fluid model\cite{Omotani:2015un} by considering the balance of electrical currents through an isolated filament in the SOL.  In the region of the outboard mid-plane of a tokamak, the gradient and curvature of the magnetic field act to enhance diamagnetic currents at larger radial distances for a given pressure gradient.  The cross-field pressure gradients within the filament thus lead to diamagnetic currents in the perpendicular plane that have a non-zero divergence.  Current continuity necessitates additional currents, which can take the form of perpendicular polarization currents to produce a closed circuit within the drift-plane, or parallel currents that can close through the sheath at the target.  Each of these current paths result in the formation of a broadly dipolar electrostatic potential field in the perpendicular plane, 
which through $\boldsymbol{E}\times\boldsymbol{B}$ motions corresponds to a pair of counter rotating vortices that act to advect the filament radially outwards.  The magnitude and structure of this potential field and thus the filament's detailed motion, is dependent on whether the parallel or polarization current path is dominant in closing the diamagnetic currents, which is in turn determined by the effective resistances of each path.   

Much of the work to date concerning the theory and simulation of filament motion has concentrated on the case in which the resistance to currents traveling through the sheath, $\Gamma_{sheath}$, is much greater than the total resistance to parallel currents through the bulk SOL plasma, $\Gamma_{\parallel}$, so that resistance of the entire parallel current path is dominated by $\Gamma_{sheath}$.  In this limit, it has been shown that the perpendicular length scale of the filament perturbation, $\delta_\perp$, plays an important role in determining whether the parallel or polarization current path is dominant in closing the diamagnetic current drive, and thus the filament's motions \cite{Yu:2003eoa, Theiler:2009fh}.  Estimates of how a filament's radial velocity scales with this parameter have been derived.  For filaments much smaller than a critical length, $\delta_\perp \ll \delta_{*0}$, where $\delta_{*0}$ is defined in Equation \eqref{eq:delta_*0}, the polarization current path is dominant and the filament's velocity is estimated to scale like $\sim {\delta_\perp}^{1/2}$.  Such filaments are said to be in the \textit{inertial} regime or \textit{inertially limited}.  On the other hand, for filaments much larger than this critical length, $\delta_\perp \gg \delta_{*0}$, parallel currents traveling through the sheath to close at the target become dominant and the velocity is estimated to scale like $\sim {\delta_\perp}^{-2}$.  In this case, the filaments are described to be in the \textit{sheath current} regime or \textit{sheath current limited}.  These asymptotic regimes have been observed in two dimensional \cite{Yu:2003eoa, Omotani:2015un}, and more recently three dimensional, simulations \cite{Easy:2014eb, Halpern:2014ip}. 

There exist a number of mechanisms however by which $\Gamma_{\parallel}$ may significantly increase, such that the total resistance of the entire parallel current path is no longer dominated by $\Gamma_{sheath}$.  One such mechanism is an increased parallel connection length to the target, which can be achieved in a larger device or by use of a Super-X divertor \cite{Katramados:2011dk}.  Moreover, decreased temperatures within the SOL will lead to an enhanced electron-ion collisionality and thus a higher parallel resistivity.  In a conduction limited SOL this effect can become very strong particularly in the divertor region, as the temperature downstream at the target can become much cooler than upstream at the mid-plane.  Moreover, if the temperature at the target becomes sufficiently low ($T_e <$1eV), volume recombination becomes strong and divertor detachment will occur\cite{Stangeby:2008ta}, meaning that a cloud of neutrals forms between the plasma and the target.  Once the ionization fraction is sufficiently small, electron-neutral collisions are comparable with electron-ion collisions\cite{Goldston:1995wz}, and the resistivity has a component proportional to the ratio between neutral and electron densities\cite{Inan:2010um} (see appendix).  In the limit of zero ionization, the resistance of the neutral gas in front of the targets is effectively infinite.  

The effect of parallel resistivity was considered using a two region model in Reference \citenum{Myra:2006ff}.  In the absence of magnetic geometry effects which were also considered, the work's predictions are equivalent to the aforementioned inertial and sheath current regimes when $\Gamma_{sheath} \gg \Gamma_{\parallel}$ (although they are referred to as the \textit{resistive ballooning} and \textit{sheath connected interchange} regimes respectively).  In the case in which $\Gamma_{\parallel} \geq \Gamma_{sheath}$, the inertial regime is expected to continue for the smallest $\delta_\perp$, as the regime does not involve parallel currents.  For larger $\delta_\perp$ filaments in this case however, what is described in this work as a \textit{resistive sheath current} regime is predicted (Reference \citenum{Myra:2006ff} uses the term \textit{resistive X-point} regime), in which the diamagnetic currents are closed through the sheath, but the parallel resistance of the plasma, rather than that of the sheath is expected to dominate in determining the electrostatic potential within the filament, and thus its radial velocity.  The radial velocity within this regime is predicted to scale like $\sim \Gamma_{\parallel} {\delta_\perp}^{-2}$.  This two region model also predicts that the critical $\delta_\perp$ at which filaments transition from the inertial to the resistive sheath current regime scales like $\delta_* \sim \Gamma_{\parallel}^{2/5}$.  Therefore if the plasma's parallel resistivity is sufficiently high, then effectively all filaments of a physically realistic $\delta_\perp$ will be in an inertial regime.  

It is this extreme limiting case that the two dimensional ESEL model \cite{Fundamenski:2007gk} considers by neglecting the influence of parallel currents entirely.  Turbulence simulations using this model have been successful in reproducing experimental measurements of SOL profiles and turbulence statistics from a variety of experimental devices \cite{Fundamenski:2007gk, Garcia:2005ki, Militello:2013cl}.  The absence of parallel current effects within this model mean that it only reproduces the inertial regime.  Isolated filament simulations using the ESEL model therefore did not find good agreement with three dimensional simulations in Reference \citenum{Easy:2014eb}, which used parameters such that $\Gamma_{sheath} \gg \Gamma_{\parallel}$ and so also exhibited the sheath current regime.  It was however suggested that inclusion of additional physics such as an enhanced collisionalities or divertor detachment within the three dimensional model would make its results more comparable to those obtained using the ESEL model. 

The three dimensional simulations presented in this work therefore investigate the effect of enhanced resistivity, particularly in the region nearest to the targets, on the current balance within filaments and hence their radial velocity.  The remainder of this paper is organized as follows.  Section \ref{sec:eqns} provides an outline of the physical model used in this work, before Section \ref{sec:sim_implementation} describes the numerics, boundary conditions and initialization of the simulations.  The results in Section \ref{sec:results} are split into three parts.  First Section \ref{sec:reference} describes the dynamics of a low resistivity case that is used as a reference case which the enhanced resistivity simulations are compared against.  Next, the effect of enhancing the resistivity only in the region nearest the target is investigated in Section \ref{sec:sheath_loc}, before the Section \ref{sec:constant} demonstrates the effect of resistivity enhanced uniformly throughout the entire domain.  Finally the main conclusions of this work are summarized in Section \ref{sec:conclusions}.  

\section{Physical Model} \label{sec:eqns}
The simulations presented in this paper have been obtained using the same physical model as in Reference \citenum{Easy:2014eb}, which is an electrostatic drift-fluid model that assumes singly charged cold ions and isothermal electrons.  It is acknowledged that the assumption of cold ions is poorly justified in the SOL, where typically $T_i \geq T_e$ \cite{Elmore:2012cw}, but it is used in this work for simplicity.  

The effects of magnetic geometry have been neglected by employing a slab geometry with uniform magnetic field $\boldsymbol{B} = B\hat{z}$ to represent the SOL, with the effects of magnetic curvature and gradients included through additional terms in the evolution equations.  The radial coordinate in this geometry is represented by the $x$ coordinate, whilst $y$ corresponds to an effective poloidal coordinate.  Throughout this work a Bohm normalization is used, with time and length scales normalized to the ion gyro-frequency, $\Omega_i = eB/m_i$, and the hybrid gyro-radius $\rho_s = c_s/\Omega_i$ respectively, whilst the electrostatic potential, $\phi$, has been normalized to $T_e/e$. Here $e$ is the elementary unit charge, $m_i$ is the ion mass, $c_s = \sqrt{T_e/m_e}$ is the sound speed, $T_e$ is the electron temperature in Joules and $m_e$ is the mass of an electron.  In addition, the plasma density has been normalized to a characteristic SOL plasma density, $n_0$.  The resulting dimensionless evolution equations for plasma density, $n$, vorticity, $\Omega = \nabla_\perp^2\phi$, parallel ion velocity, $U$, and parallel electron velocity, $V$, are:
\begin{align} 
\dfrac{d\Omega}{dt} 
&= -U\nabla_\parallel\Omega 
+ \dfrac{1}{n}\nabla_\parallel J_\parallel 
- \dfrac{g}{n} \dfrac{\partial n}{\partial y}
+ \mu_i\nabla_\perp^2\Omega, 
\label{eq:vort}
\\
\dfrac{dn}{dt}
& = - \nabla_\parallel\left( n V\right) 
+ ng\dfrac{\partial \phi}{\partial y} 
- g \dfrac{\partial n}{\partial y} 
+ D_n\nabla_\perp^2 n 
+ S_n,
\label{eq:n}
\\
\dfrac{dU}{d t} 
&= -U\nabla_\parallel U
- \nabla_\parallel\phi
- \dfrac{\nu_{\parallel}}{\mu}J_\parallel
- \dfrac{S_nU}{n},
\label{eq:U}
\\
\dfrac{dV}{dt} 
&= - V\nabla_\parallel V
+ \mu\nabla_\parallel\phi
- \dfrac{\mu}{n}\nabla_\parallel n
+ \nu_{\parallel}J_\parallel
- \dfrac{S_nV}{n}.
\label{eq:V} 
\end{align}
Here, $\frac{d}{dt} = (\frac{\partial}{\partial t} + \boldsymbol{\hat{z}}\times\nabla\phi\cdot\nabla)$, $J_\parallel = n(U-V)$ is the normalized parallel current density, $S_n$ is a particle source, $\mu = m_i/m_e$ is the ratio of ion to electron masses, $D_n$ is the normalized particle perpendicular diffusivity, $\mu_i$ is the normalized ion perpendicular viscosity, $\nu_\parallel = \nu_{ei0}/1.96\Omega_i$, $\nu_{ei0} = n_0 e^4\ln \Lambda/3m_e^{1/2}\epsilon_0^2 (2\pi T_e)^{3/2}$ is the electron-ion collision frequency, $\ln \Lambda$ is the Coulomb logarithm and $\epsilon_0$ is the permittivity of free space.  It is through the parameter $\nu_\parallel$ that the effective parallel resistivity, $\eta_\parallel = \nu_\parallel/\mu$ was controlled for the studies in this paper.

Equation \eqref{eq:vort} enforces current continuity and is simply the divergence of current density divided through by the plasma density, $\nabla\cdot\boldsymbol{J}/n$, under the Boussinesq approximation.  Whilst the second term on the Right Hand Side (RHS) of this equation is written explicitly  as $\nabla_\parallel J_\parallel/n$, the remaining terms within this equation originate from the other currents in the system as follows.  The left hand side and first term on the RHS result from the ion polarization current, $\boldsymbol{J}_{pol}$.  The electron diamagnetic current density, $\boldsymbol{J}_{dia} = \boldsymbol{\hat{b}}\times\nabla n/B$ leads to the third term on the RHS, with the last term arising from the viscous current density, $\boldsymbol{J}_{visc}$, which exists due to the presence of viscosity in the system.  In other works, $\boldsymbol{J}_{visc}$ is often included within $\boldsymbol{J}_{pol}$.  It is through the diamagnetic current that the $\nabla B$ and curvature effects force the other currents in the system.  The strength of such a drive has been represented through the dimensionless variable $g$, which at the outboard mid-plane of a tokamak can be approximated to be $g = 2\rho_s/R_c$, where $R_c$ is the dimensional radius of curvature.  Concerning the other governing equations of the model, Equations \eqref{eq:n} to \eqref{eq:V} conserve particle density and parallel momentum for each particle species with Equations \eqref{eq:U} and \eqref{eq:V} written in non conservation form.  

At the location of the entrance to the sheath in front of the target, $z = \pm L_\parallel$, where $L_\parallel = \ell_\parallel/\rho_s$ is the normalized mid-plane to target distance, the parallel velocity fields (which are normalized $c_s$) evolved by these two equations must satisfy standard sheath boundary conditions \cite{Stangeby:2008ta}:
\begin{align}
\left. U \right|_{z= L_\parallel} & \geq 1 \label{eq:Ubnd}, \\ 
\left. U \right|_{z= -L_\parallel} & \leq -1 \label{eq:Ubnd2}, \\ 
\left. V \right|_{z= \pm L_\parallel} & = \pm \exp\left(- \left. \phi \right|_{z= \pm L_\parallel} \right)\label{eq:Vbnd}. 
\end{align}
In writing equation \eqref{eq:Vbnd} in such form, $\phi$ is defined relative to the potential at the target plate wall $\phi_w$, which is given a fixed value:
\begin{equation}
\phi_w = -\ln\left[\left( \dfrac{\mu}{2\pi}\right) ^{1/2} \right] \label{eq:phi_w}.  
\end{equation}
The target is thus assumed to have zero resistance.  From Equations \eqref{eq:Ubnd} to \eqref{eq:Vbnd}, the resistance to parallel currents traveling through the sheath and closing at the target can be estimated.  Assuming the potential at the sheath entrance to be small, Equation \eqref{eq:Vbnd} can be linearized so that the current density traveling through the sheath is approximately $J_{\parallel,s} \approx n_s\phi_s$, where the subscript $s$ denotes the value of a field at the entrance to the sheath.  Two points at the sheath entrance which have a potential difference with equal magnitude but opposite sign correspond therefore to the parallel current density at the first point traveling through the sheath, closing through the target and emerging at the second point.  The potential difference between the two points is $2J_\parallel/n_s$ and by using basic circuit theory the resistance (for a unit area) to currents traveling through the sheath to close at the target is $2/n_s$.  Since this path involves the current going through the sheath twice, the effective resistance of the sheath itself is:
\begin{equation}
\Gamma_{sheath} = \dfrac{1}{n_s}.  
\end{equation}
The effective resistance of parallel currents traveling from the diamagnetic current source, assumed to be located at the mid-plane ($z = 0$), to the sheath entrance can also be calculated:
\begin{equation}
\Gamma_{\parallel} = \int\limits_0^{L_\parallel}\dfrac{\nu_\parallel}{\mu} \,\mathrm{d}z.  
\end{equation}

Using these effective resistance definitions, the theoretical estimates from References \citenum{Yu:2003eoa} and \citenum{Myra:2006ff} for the critical value of $\delta_\perp$ at which filaments transition from the inertial regime, $\delta_*$, can be summarized as: 
\begin{equation}
\delta_* \sim \left\{ 
  	\begin{array}{l l}
    	\delta_{*0}                         & \Gamma_\parallel \ll    \Gamma_{sheath} \\
    	\delta_{*0}{\Gamma_\parallel}^{2/5} & \Gamma_\parallel \gg \Gamma_{sheath}
  	\end{array} \right. ,
  	\label{eq:delta_*}
\end{equation}
where
\begin{equation}
\delta_{*0} = \left(\dfrac{gL_\parallel^2}{2}\right) ^{1/5}.   \label{eq:delta_*0}
\end{equation}
The $\Gamma_\parallel \ll \Gamma_{sheath}$ condition corresponds to the transition between the inertial and sheath current regimes, whilst the $\Gamma_\parallel \gg \Gamma_{sheath}$ condition corresponds to Reference \citenum{Myra:2006ff}'s prediction for a transition from the inertial regime to the anticipated resistive sheath current regime.  Equation \eqref{eq:delta_*0} can be derived by following the scaling arguments in References \citenum{Omotani:2015un} or \citenum{Angus:2012fp}.  It is noted from the former reference that $\delta_{*0}$ should also include an order unity correction to account for the magnitude of the density perturbation, but since this can only be determined numerically via an amplitude scan, it has been neglected here.  

\section{Simulation Implementation}
\label{sec:sim_implementation}
\subsection{Numerics}
\label{numerics}
The results within this paper were obtained using the SOL2Fluid physics module\cite{Easy:2014eb} written using the BOUT++ framework \cite{Dudson:2009ig, Dudson:2015wk}.  The time integration was carried out using a fully implicit Newton-Krylov Backwards Difference Formula (BDF) solver from the PVODE library.  All the spatial derivatives were calculated using second order accurate schemes and thus the code was second order accurate.  Specifically, the parallel advection derivatives were calculated using an upwind scheme, an Arakawa scheme \cite{Arakawa:1966gs} was used for the perpendicular $\boldsymbol{E}\times \boldsymbol{B}$ advective terms and all other derivatives were calculated using central differencing.  For numerical stability, the $U$ and $V$ fields were staggered in the parallel direction relative to the other fields.  Both the BOUT++ framework and the SOL2Fluid physics module has been successfully verified using the Method of Manufactured Solutions\cite{Salari:2000uv}.  

\subsection{Boundary Conditions}
\label{sec:BCs}
For computational efficiency, only half the parallel domain was simulated, with symmetry boundary conditions employed at the lower parallel boundary at $z = 0$.  At the upper parallel boundary, $z = L_\parallel$, Equations \eqref{eq:Ubnd} and \eqref{eq:Vbnd} were enforced on $U$ and $V$.  No boundary conditions were specified for the remaining variables $n$, $\phi$, $\Omega$ at the upper parallel boundary to avoid over constraint of the system.  In the perpendicular plane, the $y$ direction was periodic for all fields, whilst at the $x$ boundaries $\Omega$ and the gradients of $n$, $U$ and $V$ were set to zero.  The remaining perpendicular boundary condition for $\phi$ was set to obtain the 1D (variation only in the parallel direction) equilibrium fields described in the next subsection.  This was achieved by fixing the $x$ boundaries of $\phi$ to the parallel profile of its resulting equilibrium field.  However, as this boundary condition could not be determined \textit{a priori}, it was obtained by iteratively running the simulation until it achieved a steady state equilibrium and updating the $\phi$ boundary condition to its parallel profile along the center of the domain until a time invariant 1D system was produced.  

\subsection{Initialization}
\label{sec:Init}
\begin{figure}
\centering
\includegraphics[trim = 0mm 0mm 0mm 0mm, clip, width = 8.5cm]{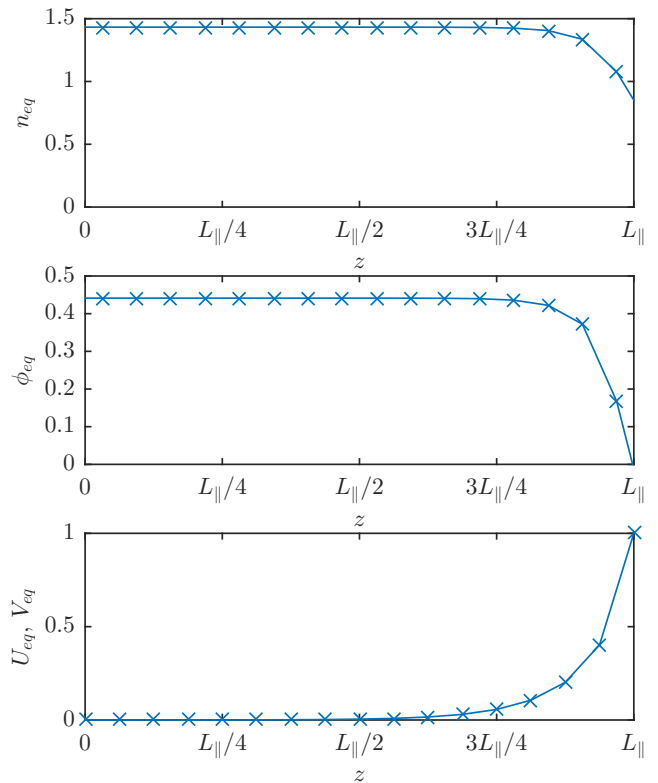}
\caption{Variation along the parallel direction of the equilibrium fields used for all of the simulations in this work.  The fields are all uniform in the perpendicular plane. }
\label{fig:equilibriums}	 
\end{figure}

In order to isolate the dynamics of single filaments from that of the background fields, a steady state equilibrium with variation only in the parallel direction was required, onto which the filament density perturbations could be seeded.  This was achieved by evolving the system until time invariant background fields were obtained using the following density source:  
\begin{equation}
\label{eq:Sn}
S_n = \dfrac{10\exp\left(10z/L_\parallel \right)}{L_\parallel\left( \exp\left(10 \right) - 1 \right)}.
\end{equation}
The source is predominantly localized in the last 10\% of the domain nearest the target, and this structure was chosen as it produces equilibrium fields with negligible parallel velocities and parallel gradients of density and potential for the majority of the domain.  The source structure can also be interpreted to loosely model a high recycling regime.  The equilibrium field profiles (denoted by the subscript $eq$) produced using this source are shown in Figure \ref{fig:equilibriums}, with the exception of $\Omega_{eq}$, which is necessarily zero.  This equilibrium has been verified against analytical results\cite{Easy:2014eb} to ensure that the equations were being solved correctly.  

The filaments were modeled within the simulations as density perturbations on top of the equilibrium density field.  It is emphasized that the simulation code evolved the quantities according to the full non linear equations, and that the equilibrium and filament perturbation were not evolved separately.  Each simulation was seeded with a single density perturbation of the form
\begin{widetext}
\begin{equation}
\label{eq:nf}
n_{f}\big| _{t = 0}  = 2 n_{eq}\big| _{z = 0} \exp\left( -\dfrac{x^2+y^2}{\delta_\perp^2}\right) \left\lbrace 1 - \tanh\left[ \dfrac{10}{L_\parallel}\left( z-\dfrac{L_\parallel}{2}\right) \right]\right\rbrace . 
\end{equation}
\end{widetext}
where $n_{f} = n - n_{eq}$, is the density perturbation of the filament.  The filaments were thus seeded as Gaussian structures in the perpendicular plane with a perpendicular length scale $\delta_\perp$.  In the parallel direction they extended approximately half the length of the domain from the mid-plane at $z=0$ to $z = L_\parallel/2$.  This structure along the field line was chosen as filaments are believed to be generated inside of the separatrix in the outboard mid-plane region\cite{Gunn:2007cx}, and have been observed in MAST to extend from X-point to X-point within the SOL \cite{Kirk:2006tt}.  Moreover, 3D filament simulations that included realistic magnetic geometry have shown that filaments initialized extending from target to target will rapidly develop parallel density gradients due to ballooning motions around the outboard mid-plane and enhanced dissipative effects downstream of the X-point region\cite{Walkden:2014tb}.  

Each simulation utilized a grid mesh of $N_x \times N_y \times N_z = 192 \times 128 \times 16$ grid points, where $N_i$ is the number of grid points in the $i$ coordinate direction, and the perpendicular domain size was scaled with $\delta_\perp$ so that the lengths of the simulation domain in the $x$ and $y$ directions were $L_x = 15\delta_\perp$ and $L_y = 10 \delta_\perp$ respectively. 

\section{Results}
\label{sec:results}
Each simulation was evolved until $t = 500$ with an output time step of $5$ for $\delta_\perp \leq 5$ and 25 for $\delta_\perp > 5$, which was sufficient to obtain a characteristic radial velocity, $v_f$, of the filament.  This characteristic velocity was defined as the first maximum that occurred of $\overline{v_x}$, which is the instantaneous radial center of mass velocity of the filament perturbation:

\begin{equation}
\overline{v_x} = \dfrac{\displaystyle{}\int\limits_{-\infty}^{\infty}\int\limits_{-\infty}^{\infty}\int\limits_{-\infty}^{\infty} n_{f}\dfrac{\partial \phi}{\partial y}\,\mathrm{d}x \,\mathrm{d}y \,\mathrm{d}z}{\displaystyle{}\int\limits_{-\infty}^{\infty}\int\limits_{-\infty}^{\infty}\int\limits_{-\infty}^{\infty} n_{f}\,\mathrm{d}x \,\mathrm{d}y \,\mathrm{d}z}.  
\end{equation}
It is noted that $\overline{v_x}$ was not monotonic, and the first maximum was selected to avoid the effects of Boltzmann spinning motions\cite{Angus:2012fp, Easy:2014eb} that could produce a second larger maximum for the smallest $\delta_\perp$ filaments. 

Since the simulations were not evolved for long enough for the density perturbation of the filament to reach the target at $z = L_\parallel$, the density at the sheath remained largely constant at its equilibrium value, $n_s \approx 0.85$ (See Figure \ref{fig:equilibriums}) and so $\Gamma_{sheath}\approx 1.2$ all of the simulations presented in this work.  

\subsection{Reference Case}
\label{sec:reference}
In order to demonstrate the effect of resistivity, the dynamics of filaments using the parameters given in Table \ref{tab:paras} are first described.  The results using these parameters, which are such that $\Gamma_\parallel \ll \Gamma_{sheath}$, will then be used as a reference case against which the simulations with higher resistivity will be compared.  The parameters are broadly relevant to the conditions found in the Mega Ampere Spherical Tokamak (MAST)\cite{Militello:2011gm} and were chosen to allow comparison with Reference \citenum{Easy:2014eb}.  The dissipative parameters $D_n$ and $\mu_i$ were specified to be two orders of magnitude smaller than neoclassical estimates for their values \cite{Fundamenski:2007gk} (and so in effect broadly classical\cite{Braginskii:1965vl} values were used) to ensure that viscous currents played a negligible role.  

\begin{table}[h]
\caption{Reference Case Parameters}
\begin{tabular}{ c | c }
  Input Parameters & Dimensionless Parameters\\
  \hline
    $\begin{aligned}[t]
	    T_e            &=40 \text{ eV} 			     \\
	    B              &=0.5 \text{ T}               \\
	    n_{0}          &=0.8\times10^{13} \text{cm\textsuperscript{-3}} \\
	    R_c            &=1.5 \text{ m}               \\
	    \ell_\parallel &=10 \text{ m}                \\
	    m_i            &=3.32 \times10^{-27} \text{ kg}\\
	    \ln \Lambda    &=13.3
    \end{aligned}$ &
    $\begin{aligned}[t]
        g                 &=2.43\times10^{-3} \\
        \nu_{\parallel}   &=2.53\times10^{-2} \\
        D_n               &=1.8\times10^{-5}  \\
        \mu_i             &=5.0\times10^{-4}  \\
        \mu               &=3646              \\
        L_\parallel       &=5500              \\              
        \delta_{*0}       &=8.2               \\
        \Gamma_\parallel  &=3.8\times10^{-2}                
        \end{aligned}$
        \label{tab:paras}
\end{tabular}
\end{table}

\begin{figure}
\centering
\includegraphics[trim = 0mm 0mm 0mm 0mm, clip, width = 8.5cm]{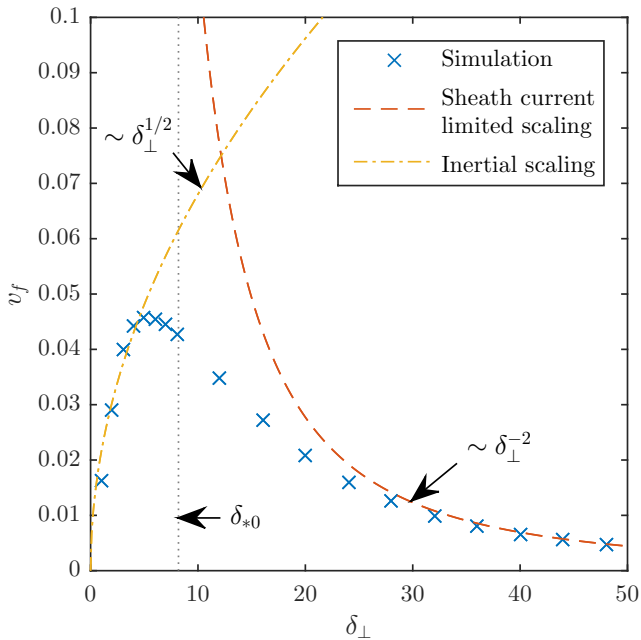}
\caption{Dependence of the characteristic radial velocity, $v_f$, on its initial perpendicular length scale $\delta_\perp$.  The analytical scaling estimates for the inertial and sheath current regimes are also plotted for comparison.}
\label{fig:ref_delta_vf}	 
\end{figure}

The dependence of $v_{f}$ on $\delta_\perp$ for the reference set of parameters is shown in Figure \ref{fig:ref_delta_vf} and it is clear that these simulations show good agreement with the $\Gamma_\parallel \ll \Gamma_{sheath}$ analytical scaling estimates discussed in Section \ref{sec:intro}, which are also plotted for comparison.  Filaments initialized with $\delta_\perp \ll \delta_{*0}$, which is plotted using a grey dotted horizontal line, produced characteristic velocities that scale like ${\delta_\perp}^{1/2}$.  On the other hand characteristic velocities proportional to ${\delta_\perp}^{-2}$ were obtained by filaments initialized with $\delta_\perp \gg \delta_{*0}$.  For reference, the peak value of $v_f = 0.46 $ corresponds to a dimensional radial velocity of 2km/s, which is consistent with experimental measurements from MAST \cite{BenAyed:2009cu}.  

\begin{figure}
\centering
\includegraphics[trim = 0mm 0mm 0mm 0mm, clip, width = 8.5cm]{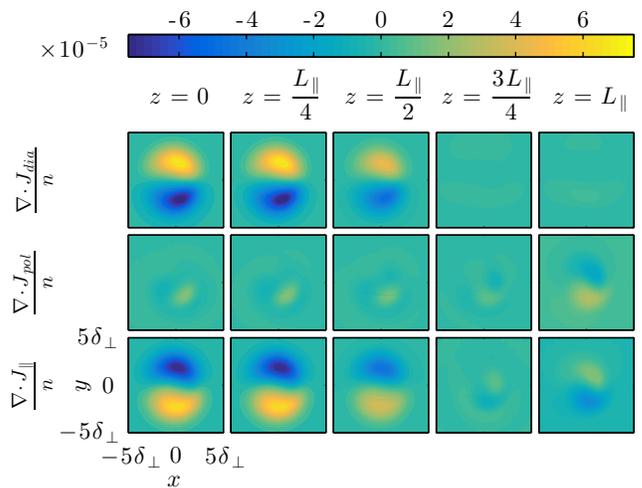}
\caption{Divergence of each current density divided by particle density, from the reference case $\delta_\perp = 28$ filament at various perpendicular planes along the field line.  The quantities are plotted at the time at which the filament's characteristic radial velocity, $v_f$, occurred.}
\label{fig:divJ_nupar1_delta28}	 
\end{figure}

\begin{figure}
\centering
\includegraphics[trim = 0mm 0mm 0mm 0mm, clip, width = 8.5cm]{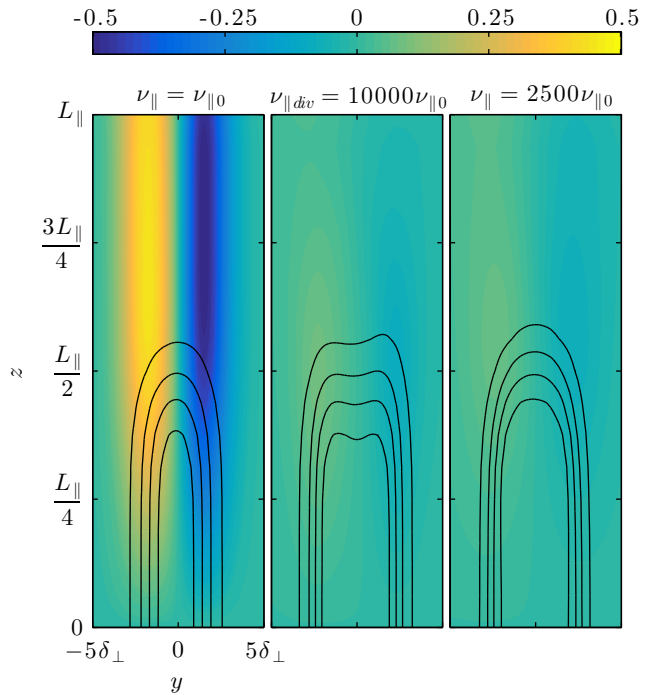}
\caption{Structure of $J_\parallel$ in a $y-z$ plane through the middle of $\delta_\perp = 28$ filaments at the time of their peak radial velocities.  The left, center and right plots respectively correspond to the reference case, the enhanced target localized resistivity $\nu_{\parallel div} = 10000\nu_{\parallel0}$ case, and the uniformly enhanced resistivity $\nu_\parallel = 2500\nu_{\parallel0}$ case filaments. Each case is plotted at the time at which the filament's characteristic radial velocity, $v_f$, occurred.}
\label{fig:J_comparison}	 
\end{figure}

It is next demonstrated that the current balance found in each regime and thus the mechanisms by which the velocities are limited are also consistent with the theoretical predictions.  The typical current balance found in large filaments, $\delta_\perp \gg \delta_{*0}$, is displayed in Figure \ref{fig:divJ_nupar1_delta28}, which plots the divergences of each of the current densities divided by $n$ (corresponding to the terms in Equation \eqref{eq:vort} as described in Section \ref{sec:eqns}) within the $\delta_\perp = 28$ filament, in various drift-planes along the field line.  The $\nabla\cdot\boldsymbol{J}_{visc}/n$ quantity has not been plotted as it is negligible.  As in all subsequent contour plots of any quantity in this work, the quantities in this figure were taken at time at which the $v_f$ occurred.  The diamagnetic current drive can be seen to exist in the region from $z = 0$ to $z=L_\parallel/2$, and is almost entirely balanced by the parallel currents.  That these parallel currents travel through the sheath and thus the filament can accurately be described as sheath current limited, is illustrated by the left plot of Figure \ref{fig:J_comparison}, which shows the structure of $J_\parallel$ in a $y-z$ plane through the middle of the filament, whose $n_f$ contours are overlaid for reference.  The parallel currents, which are of opposite sign left and right of $y = 0$, can be seen to increase in magnitude along $z$ from $J_\parallel = 0$ at $z = 0$ until the end of the density perturbation at $z \approx L_\parallel/2$.  They then remain relatively constant from there onwards as they travel through the background to the pre-sheath entrance at $z = L_\parallel$, where they can be interpreted to be traveling through the sheath to close through the target.  

\begin{figure}
\centering
\includegraphics[trim = 0mm 0mm 0mm 0mm, clip, width = 8.5cm]{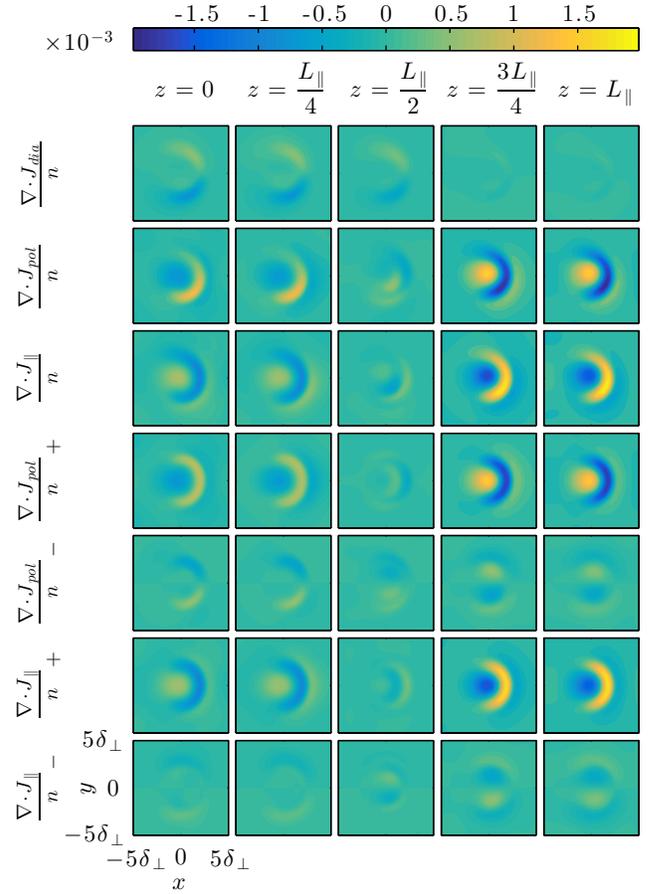}
\caption{Divergence of each current density divided by particle density, from the reference case $\delta_\perp = 5$ filament at various perpendicular planes along the field line.  The quantities are plotted at the time at which the filament's characteristic radial velocity, $v_f$, occurred.}
\label{fig:divJ_nupar1_delta5}	 
\end{figure}

A more complicated current balance is typically exhibited in small filaments, $\delta_\perp \ll \delta_{*0}$, as can be seen in Figure \ref{fig:divJ_nupar1_delta5}, which plots the divergences of current densities divided by $n$ from the $\delta_\perp = 5$ filament
.  It is not immediately obvious from the top three rows of this Figure whether the polarization or parallel currents are playing a greater role in closing the diamagnetic currents.  More clarity is provided by isolating the components of $\nabla\cdot J_{pol}/n$ and $\nabla\cdot J_{\parallel}/n$ which have even ($+$) and odd ($-$) parity in $y$ with respect to the center of mass of the filament in the poloidal direction, $y_0$.  These decompositions can be calculated for an arbitrary quantity $f$ as follows:
\begin{equation}
\label{eq:parity_decom}
f_{\pm} = \dfrac{f\left( y-y_0\right) \pm f\left( y_0-y\right)}{2},
\end{equation}
where
\begin{equation}
y_0 = \dfrac{\displaystyle{}\int\limits_{-\infty}^{\infty}\int\limits_{-\infty}^{\infty}\int\limits_{-\infty}^{\infty} n_{f} y \,\mathrm{d}x \,\mathrm{d}y \,\mathrm{d}z}{\displaystyle{}\int\limits_{-\infty}^{\infty}\int\limits_{-\infty}^{\infty}\int\limits_{-\infty}^{\infty} n_{f}\,\mathrm{d}x \,\mathrm{d}y \,\mathrm{d}z}..
\end{equation}
These decompositions are plotted in the bottom four rows of the figure.  The diamagnetic currents can be seen to be predominantly balanced by the odd component of $\boldsymbol{J}_{pol}$ and thus the filament's velocity can be said to be inertially limited.  An independent current balance is found between the even components of $J_\parallel$ and $\boldsymbol{J}_{pol}$, which arise as a result of the Boltzmann potential response to the parallel density gradients\cite{Angus:2012fp, Easy:2014eb, Walkden:2015wh} .   The odd component of $J_\parallel$ also plays an important role however, as it balances the odd component of $\boldsymbol{J}_{pol}$ in the region beyond $z=L_\parallel/2$ where the diamagnetic drive is no longer present.  This means that the odd component of $\boldsymbol{J}_{pol}$ and thus $\phi$ is broadly constant all along the field line.  

\subsection{Increased Target Localized Resistivity}
\label{sec:sheath_loc}
As discussed in Section \ref{sec:intro}, there are a number of mechanisms that may enhance the collisionality particularly in the divertor region.  Investigations were therefore carried out to determine the effect of increasing the resistivity of the plasma in this region on the dynamics of filaments by increasing $\nu_\parallel$ from its reference case value, $\nu_{\parallel0}$, in the last 25\% of the domain nearest the target: 
\begin{equation}
 \nu_\parallel = \left\{ 
  	\begin{array}{l l}
    	\nu_{\parallel0}    & \quad z \leq 3L_\parallel/4\\
    	\nu_{\parallel div} & \quad z >    3L_\parallel/4
  	\end{array} \right ..
\end{equation}
Simulations were carried out using values of $\nu_{\parallel div}$ that increased from $\nu_{\parallel0}$ by powers of ten to $10000\nu_{\parallel0}$, corresponding to values of $\Gamma_\parallel$ ranging from 0.038 to approximately 95. For comparison $\Gamma_\parallel \approx \Gamma_{sheath}$ when $\nu_{\parallel div} = 100\nu_{\parallel0}$.  An order of magnitude estimation of the electron temperature required in the divertor region to produce such enhancements in $\nu_{\parallel div}$, is provided in the appendix.  

\begin{figure}
\centering
\includegraphics[trim = 0mm 0mm 0mm 0mm, clip, width = 8.5cm]{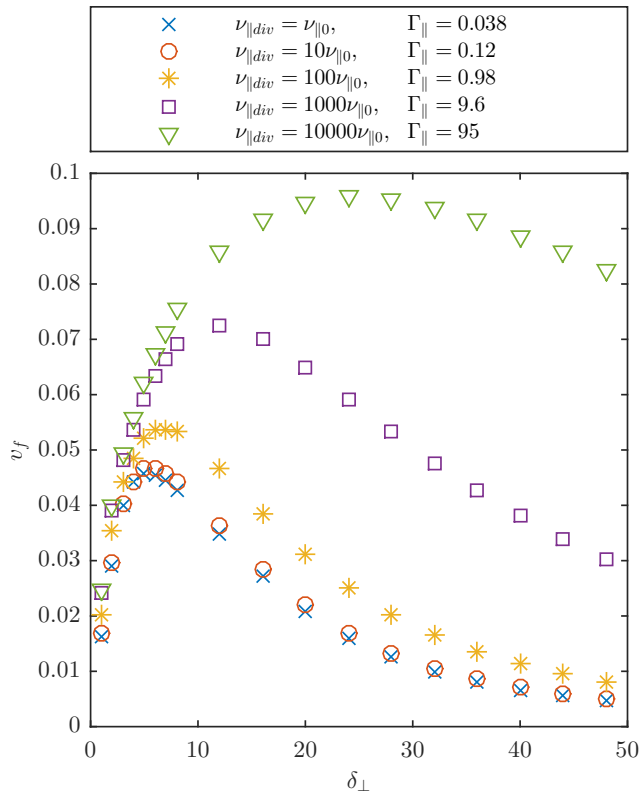}
\caption{Dependence of the characteristic radial velocity, $v_f$, on its initial perpendicular length scale $\delta_\perp$ for each of the values of $\nu_{\parallel div}$ used in the enhanced target localized resistivity scan. }
\label{fig:sheath_loc_delta_vf}	 
\end{figure}

The dependence of $v_f$ on $\delta_\perp$ for increasing values of $\nu_{\parallel div}$ can be seen in Figure \ref{fig:sheath_loc_delta_vf} and it is clear to see that increasing $\nu_{\parallel div}$ leads to enhanced radial velocities across all $\delta_\perp$, with the smallest $\delta_\perp$ experiencing a relatively modest increase in $v_f$, compared to the larger $\delta_\perp$.  This is to be expected because the smallest filaments were in the inertial regime in the reference case, meaning that parallel currents played a sub-dominant role in closing the diamagnetic currents.  On the other hand, the largest filaments were in the sheath current regime in the reference case, meaning that parallel currents were dominant in maintaining current continuity and so increasing the resistivity has a greater influence on these filaments.

\begin{figure}
\centering
\includegraphics[trim = 0mm 0mm 0mm 0mm, clip, width = 8.5cm]{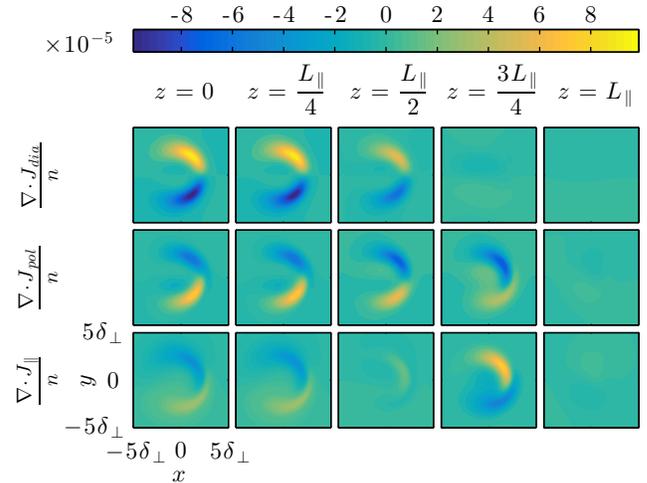}
\caption{Divergence of each current density divided by particle density, from the enhanced target localized resistivity $\delta_\perp = 28, \nu_{\parallel div} = 10000\nu_{\parallel0}$ filament at various perpendicular planes along the field line.  The quantities are plotted at the time at which the filament's characteristic radial velocity, $v_f$, occurred. }
\label{fig:divJ_nupar10000_delta28}	 
\end{figure}

One of the mechanisms by which faster velocities are produced can be understood by observing that as $\nu_{\parallel div}$ is increased, $v_f$ scales like ${\delta_\perp}^{1/2}$ up until larger values of $\delta_\perp$ and so the inertial regime is clearly extended, or equivalently $\delta_*$ is increased.  This occurs because increasing the resistivity suppresses the parallel currents and thus necessarily leads to an enhancement of the polarization currents, given the same diamagnetic current source.  The drastic reduction of $J_\parallel$ in a $\delta_\perp = 28 $ filament is evident upon comparison of the left and middle plots of Figure \ref{fig:J_comparison}, which plot this quantity in a $y-z$ plane through the center of the filament, for the reference case $\nu_{\parallel div} = \nu_{\parallel0}$ and $\nu_{\parallel div} = 10000\nu_{\parallel0}$ simulations respectively.  To demonstrate that this suppression of $J_\parallel$ affects the current balance upstream, Figure \ref{fig:divJ_nupar10000_delta28} plots the divergence of current densities divided by density, from the same simulation at the time of its peak radial velocity.  By cross comparison with Figure \ref{fig:divJ_nupar1_delta28}, it is clear that the reduction of parallel currents within the filament means that the diamagnetic current drive is predominantly balanced by enhanced polarization currents instead.  The parallel currents do still play an important role however, in that they balance $\boldsymbol{J}_{pol}$ in the region between the end of the filament perturbation and the start of the enhanced resistivity region, $L_{\parallel}/2 < z < 3L_\parallel/4$, where the diamagnetic drive is no longer present, so that $\boldsymbol{J}_{pol}$, and thus $\phi$, is approximately constant in $z$ up until the region of enhanced resistivity.  This is analogous to the role they played for $\delta_\perp \ll \delta_{*0}$ filaments in the reference case.  

Greater radial velocities were also produced at higher resistivities by the very largest $\delta_\perp$, in which parallel currents closing at the target were still the main way in which the diamagnetic currents were closed.  These filaments attained greater velocities because the resistance of the plasma was sufficient to introduce a potential difference between the downstream at the sheath entrance and further upstream in the region of the filament density perturbation.  Therefore for the same amount of current to flow into or out of the sheath, larger potentials were formed upstream at higher resistivities, which in turn correspond to faster radial velocities.  Such filaments are hereafter described to be in a \textit{resistive sheath current} regime.  

\begin{figure}
\centering
\includegraphics[trim = 0mm 0mm 0mm 0mm, clip, width = 8.5cm]{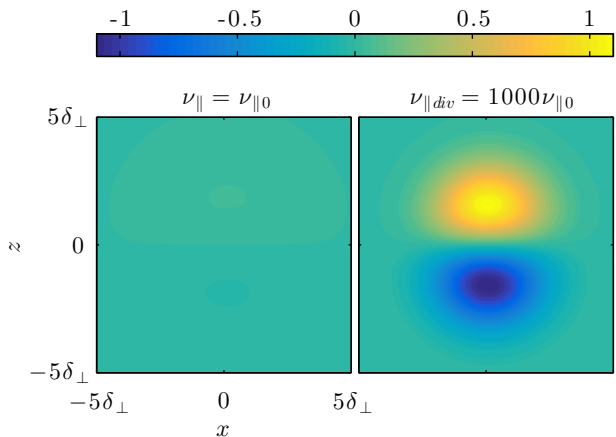}
\caption{Potential difference formed between the mid-plane and sheath, $\left. \phi_-\right|_{z = L_\parallel/2} - \left.\phi_-\right|_{z = L_\parallel}$, in $\delta_\perp = 100$ filaments, at the time of their peak radial velocity.  The left and right hand plot respectively the reference $\nu_{\parallel} = \nu_{\parallel 0}$ and enhanced resistivity $\nu_{\parallel div} = 1000 \nu_{\parallel 0}$ cases. Each quantity is plotted at the time at which the filament's characteristic radial velocity, $v_f$ occurred. } 
\label{fig:resistive_sheath_delta_phi}	 
\end{figure}

To demonstrate the potential difference formed along the parallel direction at high resistivities, it is necessary to separate it from the potential difference that is produced by the Boltzmann potential response to the parallel density gradients in the filament\cite{Angus:2012fp, Easy:2014eb, Walkden:2015wh}.  This can be achieved by isolating the component of $\phi_{-}$ according to Equation \eqref{eq:parity_decom}.  The potential difference formed at high resistivities is thus demonstrated by the right hand plot of Figure \ref{fig:resistive_sheath_delta_phi}, which plots the difference of  $\phi_-$ between the mid-plane and sheath, from a $\delta_\perp = 100, \nu_\parallel = 1000\nu_{\parallel 0}$ filament at the time of its peak radial velocity.  For comparison, the equivalent potential difference produced by a filament of the same $\delta_\perp$ using the reference case resistivity, is illustrated in the left hand plot and is clearly negligible. 
 
\begin{figure}
\centering
\includegraphics[trim = 0mm 0mm 0mm 0mm, clip, width = 8.5cm]{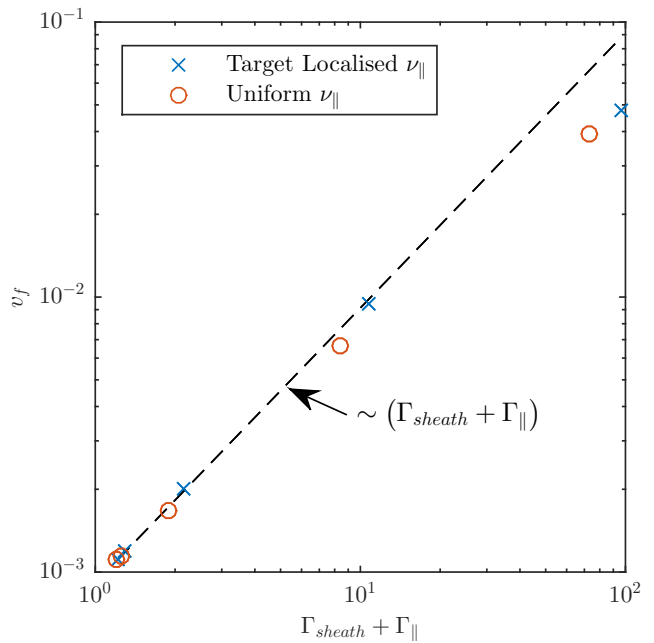}
\caption{Dependence of a $\delta_\perp = 100$ filament's characteristic radial velocity, $v_f$, on $\Gamma_{sheath} + \Gamma_\parallel$.  With the exception of the highest $\Gamma_{sheath} + \Gamma_\parallel$ point for each data series, in which polarization currents were not negligible, a linear dependence of $v_f$ on $\Gamma_{sheath} + \Gamma_\parallel$ is displayed.  $\Gamma_\parallel$ was scaled by a factor of 0.75 when plotting the uniform $\nu_\parallel$ data series, to account for the fact that the effective resistance to parallel currents traveling to the sheath from the mid-plane at $z = 0$ is approximately double that from the front of the filament at $z \approx L_\parallel/2$.  }
\label{fig:resistive_sheath_scaling}	 
\end{figure}

Despite different physical mechanisms being dominant in determining the radial velocity of the filaments in the sheath current and resistive sheath current regimes (namely sheath resistivity and plasma resistivity respectively), the radial velocities produced in both regimes are proportional to $ \left( \Gamma_{sheath} + \Gamma _\parallel \right) {\delta_\perp}^{-2}$.  The linear dependence on $\left( \Gamma_{sheath} + \Gamma _\parallel \right)$ is demonstrated in Figure \ref{fig:resistive_sheath_scaling} for $\delta_\perp = 100$ filaments.  It is noted that a value of $\Gamma_{sheath} = 1/0.85$ was used to plot this data.  The highest $\left( \Gamma_{sheath} + \Gamma _\parallel \right)$ data point deviates from this scaling because polarization currents were not negligible for this filament, and thus it is not strictly in the resistive sheath current regime.  The transition between the sheath current and resistive sheath current regime therefore occurs at the point at which the $\Gamma_\parallel \approx \Gamma_{sheath}$, as this reflects the approximate point at which the sheath and plasma resistivities play an equal role in determining the filament's velocity.  This is reflected in Figure \ref{fig:sheath_loc_delta_vf}, in that the filaments' radial velocities only noticeably deviate from their reference case values once $\Gamma_\parallel \approx \Gamma_{sheath}\approx 1$.  This transition point and the observed velocity scalings in the two regimes are consistent with the predictions of Reference \citenum{Myra:2006ff}.   

\begin{figure}
\centering
\includegraphics[trim = 0mm 0mm 0mm 0mm, clip, width = 8.5cm]{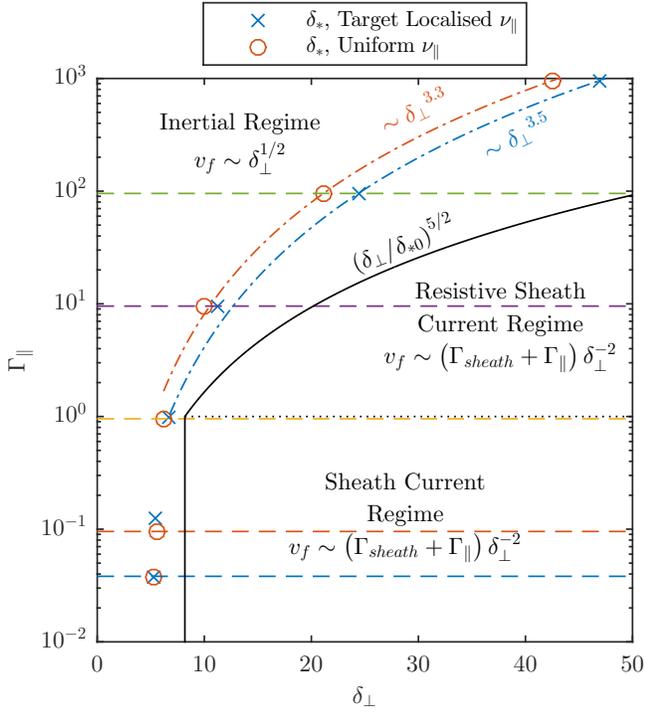}
\caption{Diagram of each filament regime location in $\Gamma_\parallel - \delta_\perp$ space.  The measured locations of the transition from the inertial regime, $\delta_*$, for different values of $\Gamma_\parallel$ are plotted using markers, whilst the analytical prediction from Reference \citenum{Myra:2006ff} for the location of $\delta_*$ is plotted using a solid black line.  The horizontal dashed lines mark the values of $\Gamma_\parallel$ corresponding to those used in the sheath localized and uniform resistivity scans, with the colors of the lines matching the colors of the markers used for their associated dataset in Figures \ref{fig:sheath_loc_delta_vf} or \ref{fig:constant_delta_vf}.}  
\label{fig:regime_plot}	 
\end{figure}

Returning to the transition from the inertial regime to the sheath current or resistive sheath current regime, $\delta_*$ was measured quantitatively from the simulation data, by defining it to be the $\delta_\perp$ at which the maximum value of $v_f$ occurred, with a cubic spline interpolation used to determine its value as accurately as possible.  The measured locations of $\delta_*$ for each $\nu_{\parallel div}$ simulated are plotted as blue crosses on the $\delta_\perp$ - $\Gamma_\parallel$ diagram in Figure \ref{fig:regime_plot} to show the observed location of the boundary of the inertial regime.  For reference, the horizontal dashed lines mark the values of $\Gamma_\parallel$ corresponding to each of the values of $\nu_{\parallel div}$ simulated, with the colors of the lines matching the colors of the markers used for their associated dataset in Figure \ref{fig:sheath_loc_delta_vf}.  Also plotted using a dotted line is the location of the transition between the sheath current and resistive sheath current regimes.  Moreover, the analytical estimates for $\delta_*$ from References \citenum{Yu:2003eoa} and \citenum{Myra:2006ff}, which are stated in Equation \eqref{eq:delta_*}, are plotted using solid black lines.  For $\Gamma_\parallel \leq 1$, the simulations' $\delta_*$ remains constant around $\delta_{*0}$ and is insensitive to $\Gamma_\parallel$ and thus good agreement is found with the analytical predictions.  For $\Gamma_\parallel > 1$, qualitative agreement is found with Reference \citenum{Myra:2006ff}'s prediction in that $\delta_*$ increases as $\Gamma_\parallel$ rises.  More quantitatively however, the observed power law dependence in this region, $\delta_* \sim  {\Gamma_\parallel}^{1/3.5} \approx{\Gamma_\parallel}^{0.28}$, obtained from the two highest $\Gamma_\parallel$ data points and plotted using a blue dash-dot line, can be seen to have a weaker scaling than the $\delta_*\sim {\Gamma_\parallel}^{2/5}$ scaling predicted in Reference \citenum{Myra:2006ff}. 

\subsection{Increased Uniform Resistivity}
\label{sec:constant}
The effect of increasing the resistivity of the plasma uniformly in the SOL, rather than just in the region nearest to the targets, was also investigated to compare and contrast against the results in the preceding subsection.  To enable a direct comparison, $\nu_\parallel$ was increased by factors such that the values of $\Gamma_\parallel$ were approximately equal to those used in the sheath localized resistivity simulations.

\begin{figure}
\centering
\includegraphics[trim = 0mm 0mm 0mm 0mm, clip, width = 8.5cm]{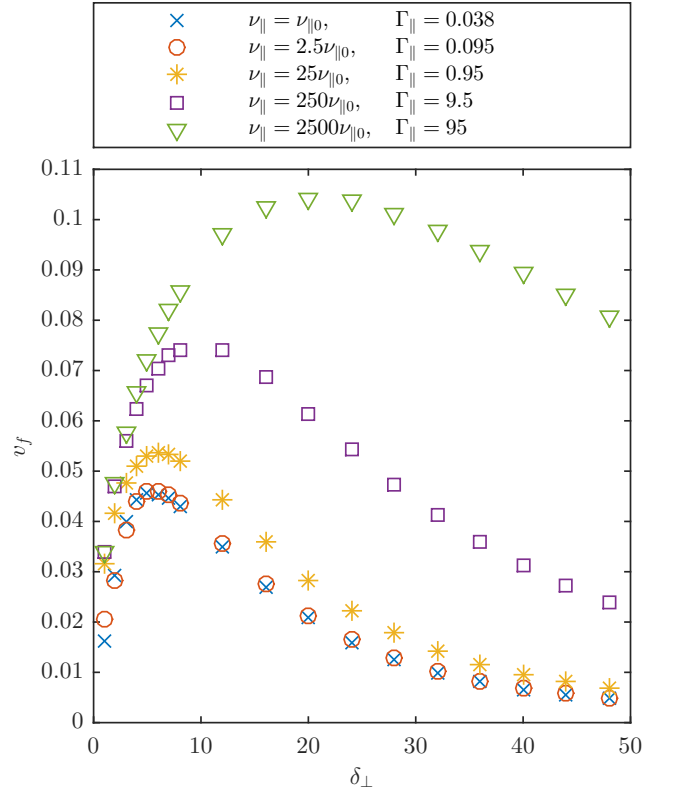}
\caption{Dependence of the characteristic radial velocity, $v_f$, on its initial perpendicular length scale $\delta_\perp$ for each of the values of $\nu_\parallel$ used in the uniformly enhanced resistivity scan. }
\label{fig:constant_delta_vf}	 
\end{figure}

In terms of the effect on $v_f$, for approximately the same value of $\Gamma_\parallel$, increasing the resistivity uniformly throughout the domain produced very similar results to increasing it only in the last 25\% nearest the target, as can be observed by comparing Figure \ref{fig:constant_delta_vf} to Figure \ref{fig:sheath_loc_delta_vf}.  

\begin{figure}
\centering
\includegraphics[trim = 0mm 0mm 0mm 0mm, clip, width = 8.5cm]{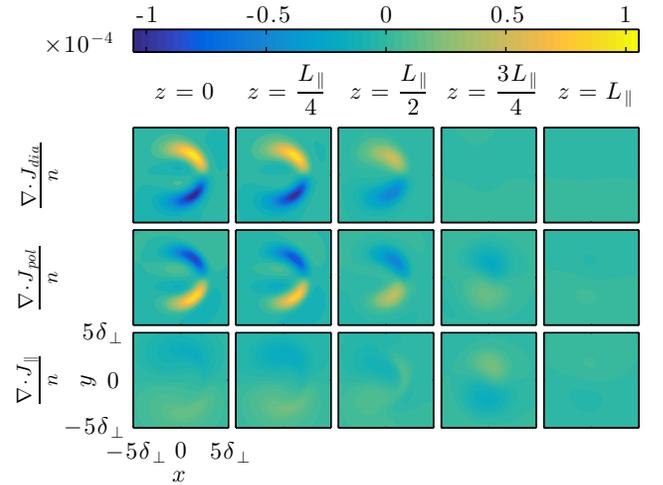}
\caption{Divergence of each current density divided by density, from the uniformly enhanced resistivity case $\delta_\perp = 28, \nu_\parallel = 2500 \nu_{\parallel0}$ filament at various perpendicular planes along the field line.  The quantities are plotted at the time at which the filament's characteristic radial velocity, $v_f$ occurred. }
\label{fig:divJ_nupar2500_delta28}	 
\end{figure}

However, the actual balance of currents produced throughout the filament is subtly different.  This can be seen by cross comparison between Figure \ref{fig:divJ_nupar2500_delta28}, which plots the divergence of current densities divided by $n$ from the $\delta_\perp = 28$, $\nu_\parallel = 2500\nu_{\parallel0}$, $\Gamma_\parallel = 95$ filament, to Figure \ref{fig:divJ_nupar10000_delta28}.  Whilst in both cases the polarization current path is dominant in closing the diamagnetic current drive, the parallel currents can be seen to play less of a role in the uniform resistivity compared to the target localized resistivity case.  This has two key effects.  The first is that for comparable $\Gamma_\parallel$, slightly higher radial velocities are attained in the inertial regime.  For example, in Figure \ref{fig:constant_delta_vf}, the $\nu_{\parallel} = 2500\nu_{\parallel 0}$ data series consistently attains higher radial velocities than $\nu_{\parallel div} = 10000\nu_{\parallel 0}$ data series in Figure \ref{fig:sheath_loc_delta_vf} for $\delta_\perp \lesssim 30$.  

\begin{figure}
\centering
\includegraphics[trim = 0mm 0mm 0mm 0mm, clip, width = 8.5cm]{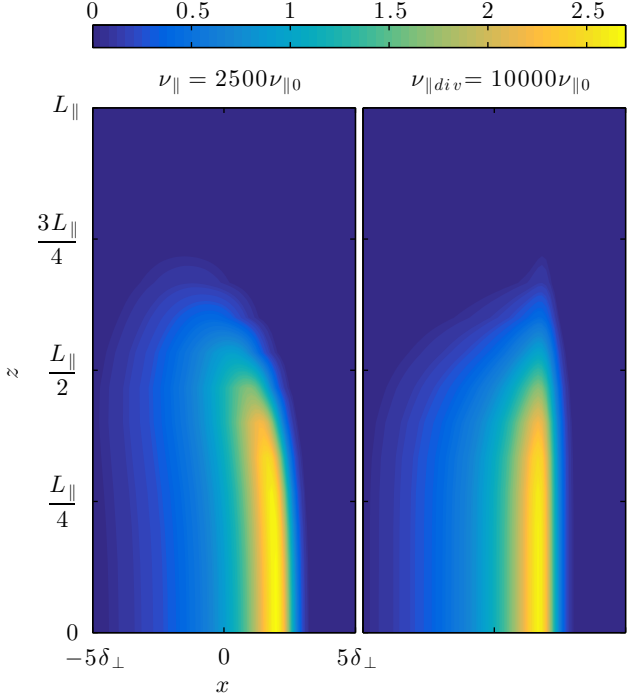}
\caption{Comparison of the structure of the density perturbation, $n_f$, in an $x-z$ plane through the middle of filaments using enhanced target localized resistivity $\nu_{\parallel div} = 10000 \nu_{\parallel0}$ and uniformly enhanced resistivity $\nu_\parallel = 2500 \nu_{\parallel0}$.  Both filaments were initialized with $\delta_\perp = 12$ and are shown at the time at which their characteristic radial velocities occurred.  The values of resistivity are such that $\Gamma_\parallel$ is approximately equal in both cases.  }
\label{fig:ballooning_n_plot}	 
\end{figure}

\begin{figure}
\centering
\includegraphics[trim = 0mm 0mm 0mm 0mm, clip, width = 8.5cm]{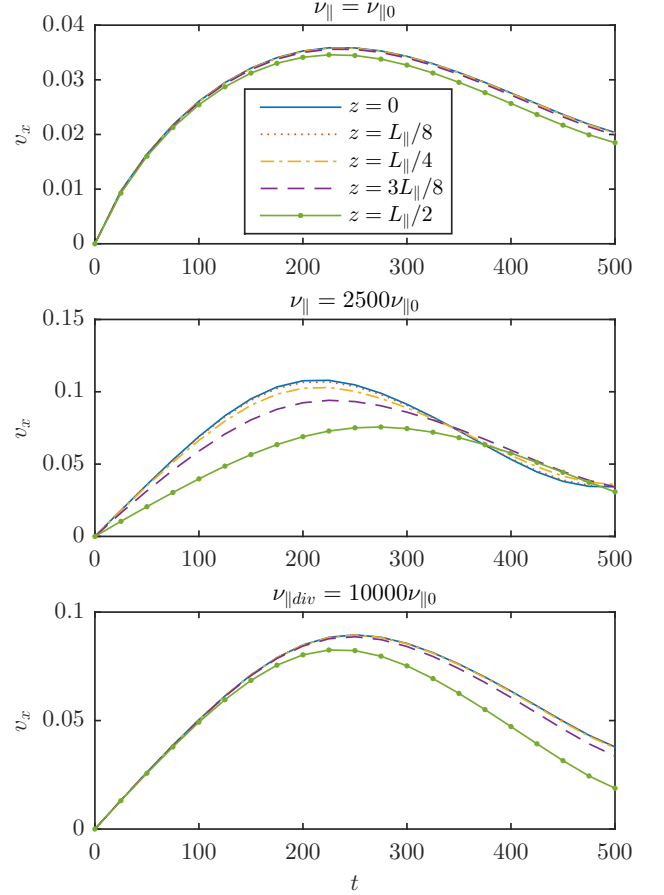}
\caption{Drift plane radial velocity, $v_x$, plotted against time at various positions along the field line, for $\delta_\perp = 12$ filaments.  The top, middle and bottom plots respectively correspond to the reference resistivity case, the uniformly enhanced resistivity $\nu_{\parallel} = 2500 \nu_{\parallel 0}$ case and the enhanced sheath localized resistivity $\nu_{\parallel div} = 10000\nu_{\parallel 0}$ case. A differential radial velocity along the field line is exhibited in the $\nu_{\parallel} = 2500 \nu_{\parallel 0}$ case.}
\label{fig:ballooning_vf_plot}	 
\end{figure}

The second effect is that a differential radial velocity along the parallel direction is produced, in that the filament moves faster in the radial direction at $z = 0$ than it does at $z = L_\parallel$.  This behavior is demonstrated by the left hand plot of Figure \ref{fig:ballooning_n_plot}, which plots $n_f$ in a $x-z$ plane through the middle of the $\delta_\perp = 12$, $\nu_\parallel = 2500 \nu_{\parallel 0}$ filament at the time at which its characteristic velocity, $v_f$, occurs.  For comparison the same quantity from the target localized resistivity $\nu_{\parallel div} = 10000\nu_{\parallel 0}$,  $\delta_\perp = 12$ simulation is plotted in the right hand plot of the same figure.  By introducing the drift plane radial velocity, 
\begin{equation}
v_x\left(z \right) = \dfrac{\displaystyle{}\int\limits_{-\infty}^{\infty}\int\limits_{-\infty}^{\infty} n_{f}\dfrac{\partial \phi}{\partial y}\,\mathrm{d}x \,\mathrm{d}y}{\displaystyle{}\int\limits_{-\infty}^{\infty}\int\limits_{-\infty}^{\infty} n_{f}\,\mathrm{d}x \,\mathrm{d}y \,},  
\end{equation}
the extent to which the radial velocity varies along the field line can be quantitatively assessed.  Figure \ref{fig:ballooning_vf_plot} plots this quantity against time at various positions along the field line in the region of density perturbation for the $\delta_\perp = 12$ simulations shown in Figure \ref{fig:ballooning_n_plot} in addition to the reference case $\delta_\perp = 12$ simulation.  It can be seen that in the uniformly enhanced resistivity case, the radial velocity at $z = L_\parallel/2$ is approximately a third slower than at $z = 0$ for most of the simulation.  

The differential radial velocity along the field line occurs in the enhanced uniform resistivity case because the parallel currents, which are suppressed throughout the domain, are not able to balance the polarization currents in the region $L_\parallel/2 < z <3L_\parallel/4$, where the diamagnetic currents are reduced or negligible.  The polarization currents therefore can only develop to match the local diamagnetic current drive, meaning that they are not constant along $z$.  Consequently $\phi$ and $v_x$ develop larger values at $z = 0$ than at $z = L_\parallel/2$, where the density gradients are smaller.  If the resistivity is high enough, $\phi$ is determined locally on each drift plane and the dynamics of the filament are effectively decoupled along the field line.  In contrast, in the enhanced target localized resistivity case, the parallel currents are able to fulfill the role of balancing the polarization currents in the region where the diamagnetic currents are absent, and so $\boldsymbol{J}_{pol}$, $\phi$ and $v_x$ are approximately constant from $z = 0$ to $z = 3L_\parallel/4$.  

Regarding the behavior of $\delta_*$ under uniformly increased resistivity, a very similar trend was displayed to what was found using an enhanced target localized resistivity.  This can be observed in Figure \ref{fig:regime_plot}, which plots using red circles the measured locations of $\delta_*$ from this uniform resistivity series of simulations.  The measured power law dependence, $\delta_* \sim {\Gamma_\parallel}^{1/3.3}\approx{\Gamma_\parallel}^{0.30}$ in the region $\Gamma_\parallel > 1$ is approximately the same as in the sheath localized resistivity case and is again weaker than Reference \citenum{Myra:2006ff}'s prediction.

\section{Conclusions}
\label{sec:conclusions}
In this work the influence of enhanced parallel resistivity on the dynamics of SOL filaments has been studied using 3D simulations.  Motivated by the expectation of lower temperatures, high neutral densities and possibility of detachment in the divertor region, the resistivity was increased only in the last quarter of the domain nearest the targets.  Increasing the resistivity lead to a suppression of parallel currents, a corresponding enhancement of polarization currents, and the development of a potential difference along the field line.  These intrinsically 3D effects meant that filaments attained higher radial velocities at enhanced resistivities.  In particular, filaments with a large perpendicular length scale, $\delta_\perp$, experienced the greatest increase in radial velocity, because at low resistivity these filaments were sheath current limited, meaning that their diamagnetic currents were predominantly closed via parallel currents.  In contrast, polarization currents were dominant in ensuring current continuity for the smallest $\delta_\perp$ filaments at low resistivities and so these filaments only experienced a modest increase in their radial velocities at higher resistivities. 

More specifically, one mechanism by which greater radial velocities were produced at higher resistivities, is that filaments that were in the sheath current regime transitioned into the inertial regime and so velocities scaled like ${\delta_\perp}^{1/2}$ up until larger values of $\delta_\perp$.  The critical $\delta_\perp$ at which this transition occurs, $\delta_*$, thus increased with resistivity and its dependence on the total resistance to parallel currents between the mid-plane and sheath entrance, $\Gamma_\parallel$, was measured to be approximately $\delta_* \propto {\Gamma_\parallel ^{0.3}}$, which is marginally weaker than that predicted by Reference \citenum{Myra:2006ff}.  Enhanced radial velocities were also observed in filaments that were sufficiently large ($\delta_\perp \gg \delta_*$), such that the parallel current path remained dominant over the polarization current path even at enhanced resistivities.  The mechanism for these resistive sheath current regime filaments was that the resistance of the plasma was sufficient to introduce a potential difference between downstream at the sheath entrance and further upstream in the region of the filament density perturbation, such that for the same amount of current to flow into the sheath, larger potentials were formed upstream at higher resistivity, corresponding to faster radial velocities.  

Investigations were also carried out in which the resistivity was increased uniformly throughout the domain.  For the same value of $\Gamma_\parallel$, marginally faster radial velocities were produced in the uniform resistivity case, as the parallel currents were more effectively suppressed.  The biggest difference with respect to the target localized resistivity simulations however was that filaments exhibited a differential radial velocity along the field line in the uniformly enhanced resistivity case, moving radially faster at the mid-plane than further downstream.  This demonstrated that enhanced resistivity can decouple the dynamics of filaments along the parallel direction.  

A limitation of the simulations presented in this paper is that since the model used assumes isothermal electrons and neglects neutral physics, the target localized resistivity was arbitrarily increased rather than self-consistently calculated.  Moreover the simulations have neglected the influence of enhanced magnetic shear around the X-point region, which could provide alternative current paths for the diamagnetic currents to be closed.  Furthermore, the assumption of cold ions is poorly justified in the SOL, where typically $T_i \geq
T_e$ \cite{Elmore:2012cw}.  Previous 2D gyrofluid model simulations\cite{Madsen:2011hu, Wiesenberger:2014em} have demonstrated that the inclusion of finite Larmor radius effects increase the coherence of filaments as they move radially outwards and induces the filament to move in the poloidal direction.  Therefore the inclusion of electron temperature dynamics, hot ions, neutral physics and magnetic geometry effects to this model would be useful additions for future research.  Nevertheless, this work has demonstrated the mechanisms by which enhanced divertor resistivities may produce faster radial filament velocities.  

\begin{acknowledgments}
This work has been part-funded by the RCUK Energy Programme [grant number EP/I501045]. To obtain further information on the data and models underlying this paper please contact PublicationsManager@ccfe.ac.uk.  In addition, this work was carried out also using the Plasma HEC Consortium EPSRC Grant No. EP/L000237/1 and the HELIOS supercomputer system at Computational Simulation Centre of International Fusion Energy Research Centre (IFERC-CSC), Aomori, Japan, under the Broader Approach collaboration between Euratom and Japan, implemented by Fusion for Energy and JAEA.
\end{acknowledgments}

\appendix*

\section{Estimate of the magnitude of  $\boldsymbol{\nu_{\parallel div}}$ at low divertor temperatures }
This appendix provides an order of magnitude estimate of the electron temperature in the divertor region, $T_e^{div}$, that may be required to produce the values of $\nu_{\parallel div}$ used in the target localized resistivity study in Section \ref{sec:sheath_loc}.  The definition of $\nu_\parallel$ given in Section \ref{sec:eqns} is based upon electron-ion collisions, and is such that $\nu_\parallel \propto T_e^{-3/2}$.  However, at very low temperatures, collisions between electrons and neutrals may become important in the divertor region, and so more generally, $\nu_{\parallel}$ can be defined\cite{Inan:2010um} as:
\begin{equation}
\label{eq:nu_parallel}
\nu_\parallel = \nu_\parallel^{ei} + \nu_\parallel^{en}. 
\end{equation}
Here $\nu_\parallel^{ei} = \nu_{ei0}/\left( 1.96\Omega_i\right) $ is the normalized electron-ion collisionality (given as the definition of $\nu_\parallel$ in Section \ref{sec:eqns}), and $\nu_\parallel^{en} = \nu_{en}/\left( 1.96\Omega_i n_e/n_0\right) $ is the normalized electron-neutral collisionality where $n_e$ is the density of electrons.  Furthermore $\nu_{en} = n_n\left\langle\sigma v \right\rangle$ where $n_n$ is the density of neutral atoms, $v$ is the velocity of electrons, $\sigma$ is the cross section for collisions between electrons and neutrals (and in principle is a function of $v$), whilst $\left\langle \cdot \right\rangle$ denotes averaging over all velocities in the (assumed) Maxwellian distribution function.  The densities of neutral deuterium atoms and electrons (or deuterium ions) at a given temperature were estimated using the Saha equation \cite{chen2013introduction}, alongside the assumption that $n_e + n_n = n_0$.  It is noted that the Saha equation assumes the plasma and neutral gas to be in thermal equilibrium, which may not be a valid assumption for edge plasmas.  Moreover, its use implies an equilibrium between ionization and recombination processes, which may not occur because recombination is a relatively slow process compared to the typical timescales of fluctuations in the divertor.  Values of $\sigma$ for elastic collisions between electrons and hydrogen atoms, obtained from Reference \citenum{janev2013atomic}, were used.  These calculations arguably provide a conservative estimate of the resistivity in the divertor region, as collisions with neutral particles other than deuterium, that may be present due to sputtering or impurity seeding, have not been included.  Furthermore, anomalous resistivity effects have also been neglected.  

\begin{figure}[h]
\centering
\includegraphics[trim = 0mm 0mm 0mm 0mm, clip, width = 8.5cm]{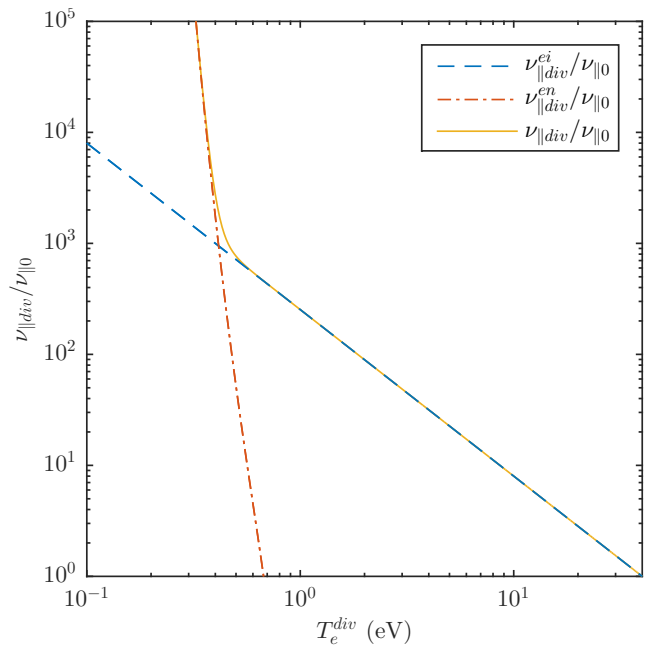}
\caption{Estimate of relative magnitude of the normalized collisionality in the divertor region, $\nu_{\parallel div}$, (and each of its constituent terms, $\nu_{\parallel div}^{ei}$ and $\nu_{\parallel div}^{en}$) compared to $\nu_{\parallel 0}$ as a function of temperature in the divertor, $T_{e}^{div}$. }
\label{fig:nu_parallel_Te}	 
\end{figure}

Figure \ref{fig:nu_parallel_Te} thus shows the estimated relative increase of the normalized collisionality in the divertor region, $\nu_{\parallel div}$ (and each of its constituent terms, $\nu_{\parallel div}^{ei}$ and $\nu_{\parallel div}^{en}$) over $\nu_{\parallel 0}$ as the temperature in the divertor, $T_{e}^{div}$ is decreased.  Moreover, for reference, the corresponding values of $T_e^{div}$ estimated to produce the values of $\nu_{\parallel div}$ including and excluding neutral collisions are given in Table \ref{tab:Te}.  These calculations indicate that at temperatures around 0.5eV, electron collisions with neutrals may dominate over electron collisions with ions.  

It is emphasized however, that the calculations presented in this Appendix are based upon a number of assumptions that may not be well justified in the SOL near the sheath, and so should only be used to give an order of magnitude indication of how the resistivity in the divertor region may depend on $T_e$.  Moreover, it is important to note that $\nu_\parallel^{en}$ dominates at low temperatures because the ratio $n_n/n_e$ becomes very large ($>1000$) and so the plasma is estimated to be very weakly ionized.  It is unclear whether such a weakly ionized plasma is achieved experimentally in the divertor region when detachment occurs.  

\begin{table}[h]
\caption{Estimated divertor temperature required to produce the values of $\nu_{\parallel div}$ used in Section \ref{sec:sheath_loc}.  }
  \label{tab:Te}
\begin{tabular}{ c | c | C{2.5cm} | C{2.5cm} }
  $\nu_{\parallel div}$ & $\Gamma_{\parallel}$ & $T_e^{div}$ required excluding neutral effects (eV) & $T_e^{div}$ required including neutral effects (eV)\\
  \hline
  $\nu_{\parallel0}$ & 0.038 & 40 & 40 \\
  $10\nu_{\parallel0}$ & 0.12 & 8.6 & 8.6 \\
  $100\nu_{\parallel0}$ & 0.98 & 1.9 & 1.9 \\
  $1000\nu_{\parallel0}$ & 9.6  & 0.40 & 0.45 \\
  $10000\nu_{\parallel0}$ & 95   & 0.086 & 0.37 \\
  $100000\nu_{\parallel0}$ & 950   & 0.019 & 0.32
\end{tabular}
\end{table}

\bibliography{bibliography}

\end{document}